\documentclass[a4paper,10pt]{article}
\usepackage[T1]{fontenc}
\usepackage[utf8]{inputenc}
\usepackage{lmodern}
\usepackage{graphicx}
\usepackage{booktabs}
\usepackage{multirow}
\usepackage{tabularx}
\usepackage[english]{babel}
\usepackage[english]{varioref}
\usepackage[autostyle, babel]{csquotes}
\usepackage[backend=biber]{biblatex}
\usepackage{hyperref}
\usepackage{geometry}
\usepackage{xcolor}
\usepackage{caption}
\usepackage{subfig}
\usepackage{amsmath}
\usepackage{amssymb}
\usepackage{textcomp}
\usepackage{times}
\title{Search for $K^{+}\rightarrow\pi^{+}\nu\overline{\nu}$ at NA62}
\author{Jacopo Pinzino\thanks{for the NA62 Collaboration: 
		G.~Aglieri Rinella, R.~Aliberti, F.~Ambrosino, R.~Ammendola, B.~Angelucci, 
		A.~Antonelli, G.~Anzivino, R.~Arcidiacono, I.~Azhinenko, 
		S.~Balev
% ...
}\\
INFN Sezione di Pisa, Italy\\
E-mail: jacopo.pinzino@cern.ch}

\begin{document}

\maketitle

\begin{abstract}
$K^{+}\rightarrow\pi^{+}\nu\overline{\nu}$ is one of the theoretically cleanest meson decay where to look for indirect effects of new physics complementary to LHC searches. The NA62 experiment at CERN SPS is designed to measure the branching ratio of this decay with 10\% precision. NA62 took data in pilot runs in 2014 and 2015 reaching the final designed beam intensity. The quality of 2015 data acquired, in view of the final measurement, will be presented.
\end{abstract}

\section{The NA62 experiment}
\subsection{Introduction}

The NA62 experiment is located in the CERN North Area SPS extraction site and it aims at measuring the Branching Ratio of the ultra-rare FCNC kaon decay $K^{+}\rightarrow\pi^{+} \nu\overline{\nu}$ collecting about 100 events in two years of data taking~\cite{Anelli:2005ju}.
This decay, with its neutral partner $K_{L}\rightarrow\pi^{0} \nu\overline{\nu}$, is a very useful process to study flavour physics and to obtain a stringent test of the Standard Model; the Branching Ratio of these decays can be computed with high precision~\cite{Buras:2015qea}, $BR(K^{+}\rightarrow\pi^{+} \nu\overline{\nu})(SM) = 8.4 \pm 1.0 \times 10^{-11}$ where the uncertainty is dominated by the current precision of the CKM mixing matrix input parameters.

The strong suppression of the SM contributions and the remarkable theoretical precision of the SM rate make this decay a powerful probe for possible new physics, complementary to direct searches at the LHC and potentially sensitive to much higher energy scales.
The combination of the Branching Ratio of these two decays ($K^{+}\rightarrow\pi^{+} \nu\overline{\nu}$ and $K^{0}\rightarrow\pi^{0} \nu\overline{\nu}$) allows to determine
 the $\beta$ angle of the Unitarity Triangle from K decays only and, in this way, to have a powerful test on Standard Model.

The most accurate measurement of this decay, $BR(K^{+}\rightarrow\pi^{+} \nu\overline{\nu}) = 17.3 ^{+11.5}_{-10.5} \times 10^{-11}$, was obtained by the E787 experiment and its upgrade E949 at BNL (from 1995 to 2002) which collected seven events~\cite{Artamonov:2008qb}.
NA62 aims to improving the measurement of this Branching Ratio reaching a precision of at least 10\%: the experiment is currently in data taking and the performances achieved in 2015 will be discussed.

\subsection{NA62 Experimental Setup}

NA62 uses the SPS 400 GeV/\textit{c} proton beam from the SPS in order to produce \textit{$K^{+}$} decaying in-flight.

The total beam rate at the end of the beam line is of the order of 750 MHz but kaons are about 6\% of the flux.
Downstream detectors aren't affected by this large flux because the undecayed particles remain inside the beam pipe; the integrated rate over these detectors is of the order of 10 MHz.

The downstream detectors start about 100 m after the beryllium target and are distributed along 170 m longitudinally; the fiducial region for decays extends from 100 m to 165 m after the target.
Detectors have an approximate azimuthal symmetry around the beam axis, with an inner hole to let the high flux of undecayed particles pass through without hitting the downstream detectors.

\begin{figure}
	\centering
	\includegraphics[width=0.8\textwidth]{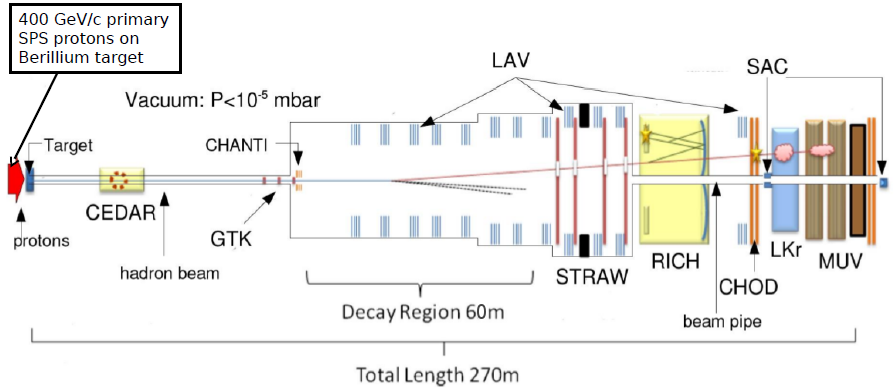}
	\caption{Longitudinal view of the NA62 experimental setup.}
	\label{fig:detectors}
\end{figure}

The NA62 experimental setup~\cite{TD}, shown in figure \ref{fig:detectors}, consists of these detectors:

\begin{itemize}
 \item The \textit{Cerenkov differential counter} (KTAG) is used to identify $K^+$ in the beam.
 It has a time resolution of about 100 ps to tag the kaon time.
 \item The Gigatracker (GTK) is composed by three silicon pixel stations placed in vacuum, with transverse dimensions which cover the beam area, and is used to measure particles direction and momentum before they enter the decay region.
 The GTK has to cope with the full beam intensity of about 750 MHz and provides a time resolution of the order 200 ps to avoid a wrong matching of a beam particle to the reconstructed decay downstream, and a resulting error in the calculation of the missing mass. Between the stations, 4 magnetic dipoles make an achromatic spectrometer for any momentum: the momentum resolution is 0.2\%, and the angular resolution for the particle direction is about 15 $\mu$rad.
 \item The CHarged ANTIcounter (CHANTI) is a set of scintillator rings that follow the last GTK station used as a veto for charged particles before they enter the decay region.
 \item A system of photon veto detectors covering a polar angle from 0 to about 50 mrad polar angle with respect to the beam direction using 12 large annular vetos (LAV) made of lead glass crystals with attached photomultipliers (PMT) and covering an angle from 8.5 to 50 mrad, a liquid krypton electromagnetic calorimeter (LKr) for angles between 1 and 8.5 mrad, an intermediate calorimeter (IRC), made of alternating layers of lead and scintillators (shashlik), to cover the ring around the beam and a small angle calorimeter (SAC) placed at the end of the beam line after a sweeping magnet and using the same shashlik technology.
 \item A magnetic spectrometer (STRAW) made of four straw tube chambers inside the vacuum tank is used to measure the position of the decay vertex, the direction and momentum of the charged secondary particle.
 The reason to operate in vacuum is to minimize the multiple scattering.
 The dipole magnet from the earlier NA48 experiment is located after the second chamber and provides a 270 MeV/\textit{c} kick in the horizontal plane, for track momentum determination.
 In the center of each chamber a region without straw let the beam particles pass undisturbed.
 \item The Ring Imaging Cerenkov (RICH) is designed to distinguish $\pi$ and $\mu$ in the momentum range between 15 and 35 GeV/\textit{c} and to measure direction and velocity of such particles.
 This detector is 17 m long, filled with Neon at atmospheric pressure and equipped with 2000 photomultipliers and has an inner beam pipe to avoid beam interactions with the gas.
 The timing resolution is of the order of 100 ps.
 \item The Charged Hodoscope (CHOD) is placed after the RICH to reduce the inefficiency in photon detection due to conversion or photo-nuclear interactions inside the material of the RICH; moreover it is used for trigger purposes.
 \item A system of muon vetoes composed of two iron-scintillator hadronic calorimeters (MUV1 and MUV2), and a plane of fast scintillators (MUV3) placed after an iron wall, gives additional power in muon vetoing and a fast trigger information.
\end{itemize}

\section{Experimental strategy}

\begin{figure}
	\centering
	\includegraphics[width=0.8\textwidth]{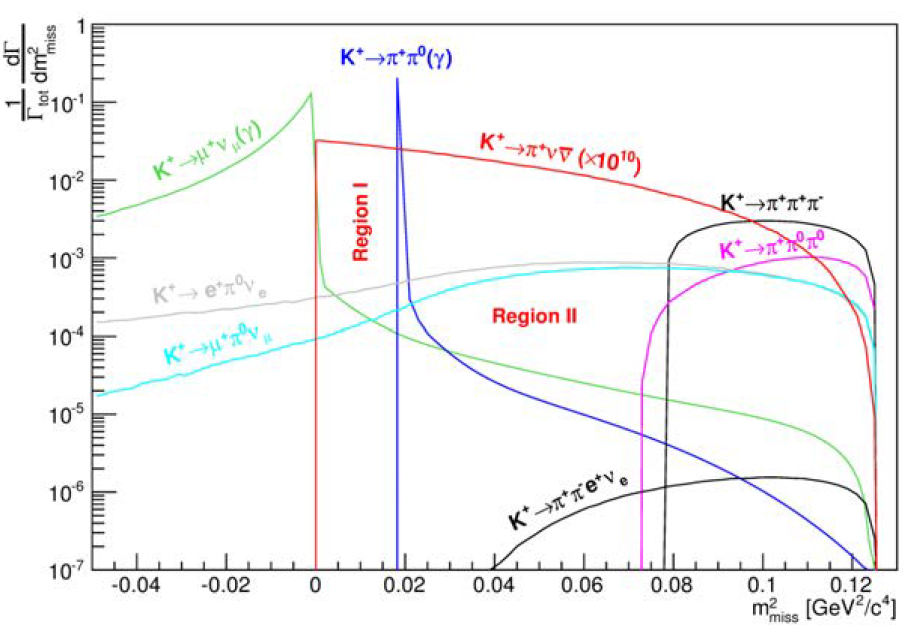}
	\caption{$m^{2}_{miss}$ distributions for signal and backgrounds from the main $ \mathrm{K^{+}} $ decay modes. The backgrounds are normalized according to their branching	fraction; the espected signal is shown multiplied by a factor $10^{10}$.}
	\label{fig:kinematic}
\end{figure}

The $K^{+}\rightarrow\pi^{+} \nu\overline{\nu}$ signature is one track in the final state matched in time with one \textit{$K^{+}$} track upstream the decay region and nothing else, because the two
neutrinos are undetectable.
Backgrounds can originate from all the kaon decays that result in a single detected charged track with no other particles, or from beam related activity.
Kinematic reconstruction is a useful rejection technique: the squared missing mass distribution of
the signal, $m^{2}_{miss} \stackrel{def}{=} (P_{K}-P_{\pi^{+}})^{2}$ (where $P_{K}$ and $P_{\pi^{+}}$ are, respectively, the 4-momenta of the kaon and the charged particles produced from kaon decay under the $\pi^{+}$ mass hypothesis), has a three body decay shape, while more than 90\% of the charged kaon decays are mostly peaking, as shown in figure \ref{fig:kinematic}.

The distribution of the  $m^{2}_{miss}$ for the signal and the main decay modes led to define two signal regions, where the main backgrounds should be limited by the
kinematic constrains, around the $K^{+}\rightarrow\pi^{+} \pi^{0}$ peak.
Semileptonc decays, radiative processes, main kaon decay modes via reconstruction tails and beam induced tracks span across these regions.
Therefore kinematic reconstruction, photon rejection, particle identification and sub-nanoseconds timing
coincidences between subdetectors must be employed to obtain the final background rejection.
A tight requirement on $P_{\pi^{+}}$ between 15 and 35 GeV/c boosts the background suppression further,
as will be shown in the next section.
Monte Carlo studies performed in past years ~\cite{Ruggiero:2013nxa} have shown that NA62 can reach the goal, exploiting multiple and almost uncorrelated techniques to suppress the main background sources.

\section{Preliminary result of the 2015 run and prospect for the  $K^{+}\rightarrow\pi^{+} \nu\overline{\nu}$ measurement}

\begin{figure}
	\centering
	{\includegraphics[width=.45\textwidth]{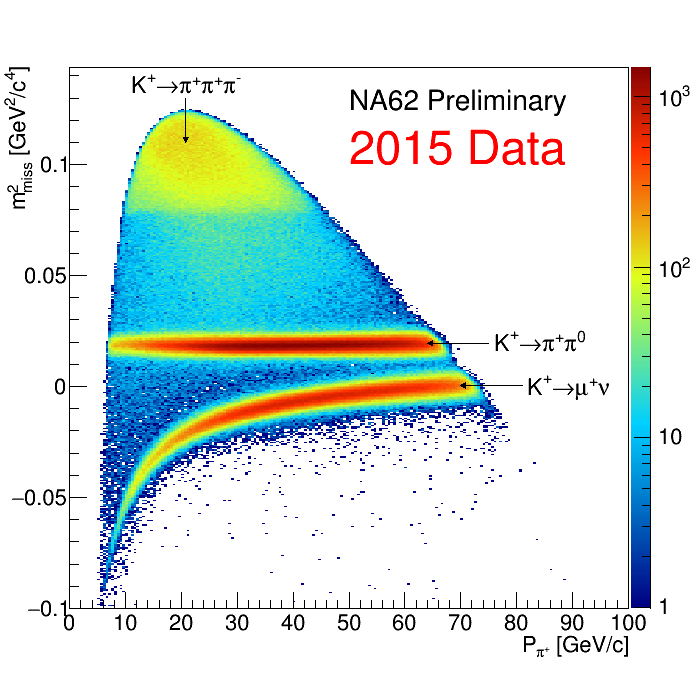}}  \quad
	{\includegraphics[width=.45\textwidth]{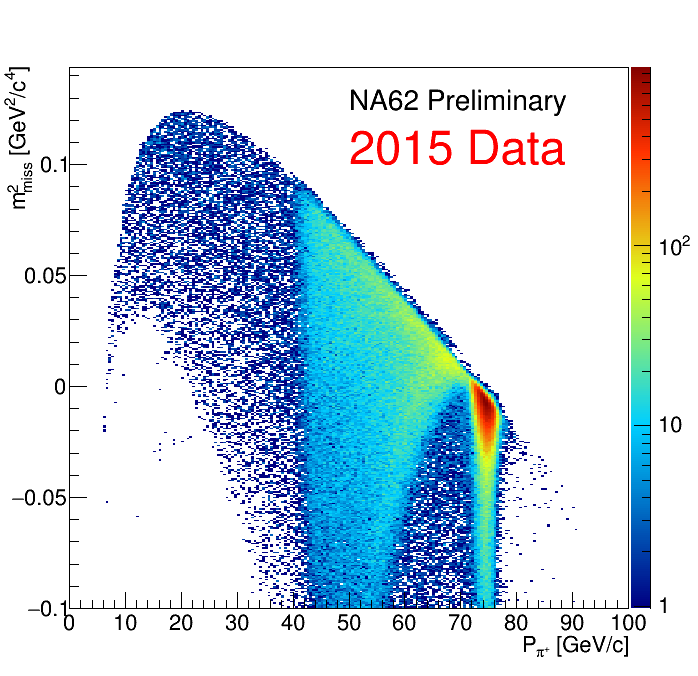}}
	\caption{$m^{2}_{miss}$ distribution under $\pi^{+}$ mass hypothesis as a function of the momentum of the track measured in the straw spectrometer after selection for single track from kaon decays (left).	Same distribution as left-side picture, but asking for single track without a positive kaon tag in	time in KTAG (right).}
	\label{fig:mmiss}
\end{figure}

The main goal of the 2015 run was to verify on data the expected detector performances, the timing,, the particle identification and the kinematic and photon rejection.
A single track selection was chosen as a preliminary step towards the $K^{+}\rightarrow\pi^{+} \nu\overline{\nu}$ measurement.
We selected tracks reconstructed in the STRAW spectrometer matching with CHOD signals and with energy depositions in calorimeters.
The CHOD signals define the track time with 200 ps resolution.
A single track event is defined by a track not forming a common vertex with any other in-time track within the decay region.
The position of the vertex is defined using using a Closest Distance of Approach (CDA) less than 1.5 cm between two tracks.
The downstream track has to match a Gigatracker track in time and space, forming a vertex in the decay region with it, in order to select events originating from kaon decays.
The Gigatracker track has to be in-time with a kaon-like signal in KTAG. 
On the right, the KTAG is used to select events related to kaons.
Time resolutions of the KTAG and GTK are found to match the design values (100 and 200 ps respectively).
The $m^{2}_{miss}$ distributions for the 2015 data, recorded at low intensity, are shown in figure \ref{fig:mmiss}: the figure on the left is done with a kaon-like signal in the KTAG, while, in the second, the KTAG is used in anti-coincidence with a Gigatracker track to select single track events not related to kaons and shows that decay from beam $\pi^{+}$, elastic scattering of beam particles in the material along the beam line (KTAG and Gigatracker stations) and inelastic scattering in the last Gigatracker station are the main sources of tracks downstream originating from beam related activity.

\begin{figure}
	\centering
	\includegraphics[width=0.8\textwidth]{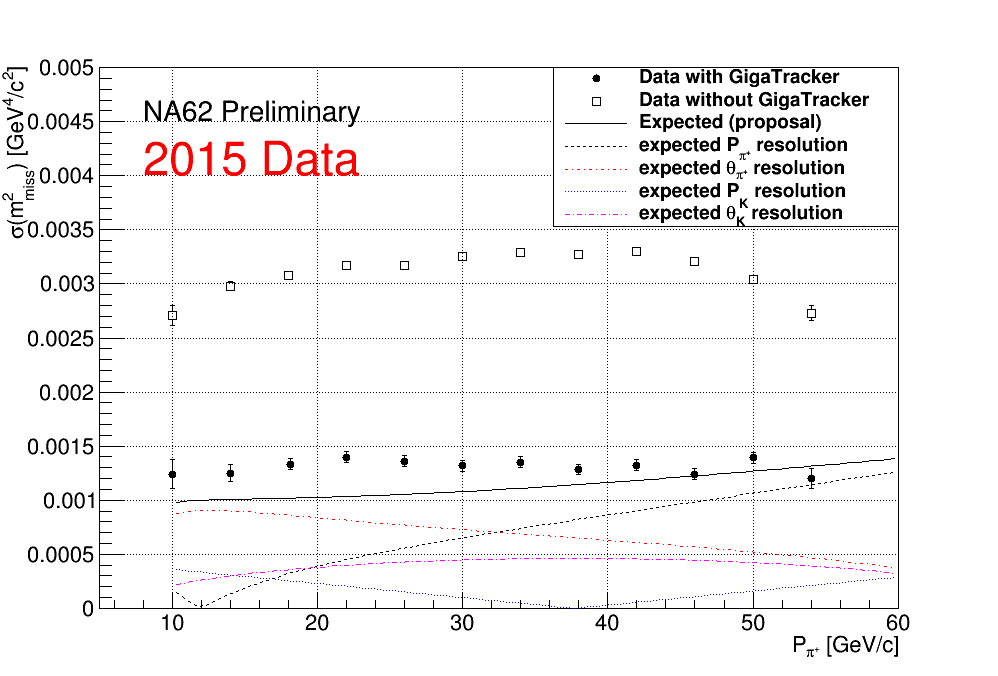}
	\caption{Resolution of the $m^{2}_{miss}$ miss vs momentum. Empty squares correspond to the values obtained with the nominal kaon momentum, black points - with the kaon momentum measured by GTK.}
	\label{fig:missmomentum}
\end{figure}

The resolution of the $m^{2}_{miss}$ is measured using the width of the $K^{+}\rightarrow\pi^{+} \pi^{0}$ peak and it is found to be $1.2 \times 10^{-3} GeV^{2}/c^{4}$ close to design value.
The resolution as a function of momentum is shown in figure \ref{fig:missmomentum}.
The resolution is a factor 3 larger if the nominal kaon momentum is taken, instead of the event by event Gigatracker measured value. 

The tracking system of NA62 is also designed to provide a rejection factor in the range of $10^{4} \div 10^{5}$ for $K^{+}\rightarrow\pi^{+} \pi^{0}$ and $K^{+}\rightarrow\mu^{+}\nu$ using $m^{2}_{miss}$ to separate signal from backgrounds, respectively.
The $K^{+}\rightarrow\pi^{+} \pi^{0}$ kinematic suppression is measured using a sub-sample of single track events from kaon decays selected by requiring the additional presence of two $\gamma$s' compatible with a $\pi^{0}$ in the LKr calorimeter. This constraint defines a sample of $K^{+}\rightarrow\pi^{+} \pi^{0}$ with negligible background even in the signal $m^{2}_{miss}$ regions, allowing the study
of the far tails of the $m^{2}_{miss}$.
The measured $K^{+}\rightarrow\pi^{+} \pi^{0}$ kinematic suppression factor is of the order of $10^{3}$.
The partial hardware Gigatracker arrangement used in 2015 mainly limits the suppression because of $m^{2}_{miss}$ tails due to beam track mis-reconstruction.

The particle identification of NA62 is designed to separate $\pi^{+}$ from $\mu^{+}$ and $e^{+}$ in order
to guarantee at least 7 order of magnitude suppression of $K^{+}\rightarrow\mu^{+}\nu$ in addition to the kinematic rejection.
The identification of secondary charged particles is done employing together RICH and calorimeters. 
The $K^{+}\rightarrow\pi^{+} \pi^{0}$ sample used for kinematic studies and a pure muon sample of $K^{+}\rightarrow\mu^{+}\nu$ were used to study the $\pi^{+}-\mu^{+}$ separation in the RICH.
The required muon contamination of 1\% was achieved with a $\pi^{+}$ ID efficiency of 80\% measured in a momentum region between 15 and 35 GeV/c.
The RICH provides also an even better separation between $\pi^{+}$ and $e^{+}$. The same $\pi^{+}$ and $\mu^{+}$ samples allow the calorimetric muon-pion separation to be investigated. Simple cut and count analysis provide a muon suppression factor within $10^{4} \div 10^{6}$ for a $\pi^{+}$ efficiency in a 90\% $\div$ 50\% range.
Several analysis techniques are under study to get the optimal separation.

The photon veto system is designed to suppress decays with photons in the final state. For photons
from $\pi^{0}$ decays the rejection power provided by LAV, LKr, IRC and SAC detectors should be at least 8 orders of magnitude.
The measured $\pi^{0}$ veto inefficiency on the 2015 data is statistically limited at $10^{6}$ (90\% CL) as an upper limit. The corresponding signal efficiency is above 90\%, being the losses mainly due to $\pi^{+}$ interactions in the RICH material producing extra clusters in LKr.
To conclude, the preliminary analysis of the low intensity 2015 data shows that NA62 is approaching the design sensitivity for measuring $K^{+}\rightarrow\pi^{+} \nu\overline{\nu}$.

\section{NA62 physics besides $K^{+}\rightarrow\pi^{+} \nu\overline{\nu}$}

The performances of the apparatus allow physics opportunities beyond the $K^{+}\rightarrow\pi^{+} \nu\overline{\nu}$ to be addressed. NA62 can significantly improve the existing limits on lepton 
flavour and number violating decays like $K^{+}\rightarrow\pi^{+} \mu^{\pm} e^{\mp}$ or $K^{+}\rightarrow\pi^{-} l^{+} l^{+}$. Experimentally $\pi^{0}$ physics can take advantage of the performances of the electromagnetic calorimeters and processes like $\pi^{0} \rightarrow invisible$ or dark photon production can be investigated. Thanks to the quality of the kinematic reconstruction, searches for heavy neutrino produced in $K^{+}\rightarrow l^{+} \nu$ decays can improve
the present sensitivity. The longitudinal scale of the apparatus open the possibility to search for
long living particles through their decays, like dark photon, heavy neutral leptons or axion-like
particles produced at the target or in beam dump configurations. NA62 is already addressing
part of the above physics program simultaneously with the  $K^{+}\rightarrow\pi^{+} \nu\overline{\nu}$ program. The full exploitation of this physics will constitute the core of the NA62 program beyond 2018.

\printbibliography

\end{document}